\documentstyle[12pt,epsf]{article}
\begin{document}

\centerline{Preprint   IBR-TH-34-05, June 1, 2005}
\centerline{To be submitted for publication in the}
\centerline{\it Proceedings of the XVIII Workshop on Hadronic Mechanics}
\centerline{University of Karlstad, Sweden,  June 20-22, 2005}

\vspace*{2cm}
\begin{center}
{\large \bf INCONSISTENCIES OF NEUTRINO AND QUARK CONJECTURES AND THEIR NEGATIVE ENVIRONMENTAL
IMPLICATIONS}

 {\bf  Ruggero Maria  Santilli}\\
{Institute for  Basic
Research\\ P. O. Box  1577, Palm Harbor,  FL 34682,  U.S.A.}\\
{ ibr@gte.net, http://www.i-b-r.org,
http://www.magnegas.com}
\end{center}

\vskip1.0cm
\begin{abstract}
By using a language as accessible to a broad audience as possible, in this note we present evidence
suggesting scientific caution prior to  final claims  that neutrinos and quarks are actual
physical particles existing in our spacetime. We review historical and recent evidence dismissing the
existence of neutrinos and quarks as physical particles, and outline recent theories representing
experimental data without their existence. We also identify the negative implications for environmental
issues of the neutrino and quark conjectures since they imply the suppression of due scientific process on
new clean energies predicted by new structure models of hadrons with massive physical constituents produced
free in spontaneous or stimulated decays.  The note
ends with the need of continuing theoretical and experimental research on neutrino and quark conjectures, but
complemented, for evident scientific democracy, accountability and societal  needs, with theoretical and
experimental studies on alternative theories without the neutrino and quark conjectures  and their
prediction of new clean energies.
\end{abstract}

\noindent {\bf 1. The litany of  directly unverifiable neutrino and quark conjectures.}

\noindent As it is well known, Rutherford [1] submitted in 1920 the conjecture that hydrogen atoms in the
core of stars are compressed into a new particle he called the {\it neutron} according to the synthesis
$$
(p^+, e^-) \rightarrow n.
\eqno(1.1)
$$

The existence of the neutron was subsequently confirm experimentally in 1932 by Chadwick [2]. However,
numerous objections were raised by the leading physicists of the time against Rutherford's
conception of the neutron as a bound state of one proton $p^+$ and one electron $e^-$.

Pauli [3] first noted that synthesis (1.1) violates  the angular
momentum conservation law because, according to quantum mechanics, a bound state of two particles with spin
1/2 (the proton and the electron) must yield a particle with integer spin and cannot yield a
particle with  spin 1/2 such as the neutron. Consequently, Pauli conjectured the
existence of a new neutral particle with spin 1/2, charge zero and no mass that is emitted in synthesis
(1.1) or in similar radioactive processes so as to verify the angular momentum conservation laws.

Fermi [4] adopted Pauli's conjecture, coined the name {\it neutrino} (meaning in Italian a "little neutron")
and presented the first comprehensive theory of the underlying interactions (called "weak"), according to
which synthesis (1.1) should be replaced with the expression
$$
(p^+, e^-) \rightarrow n + \nu,
\eqno(1.2)
$$
where $\nu$ is the neutrino, in which case the inverse reaction (the spontaneous decay of the neutron) reads
$$
n \rightarrow p^+  + e^- + \bar \nu,
\eqno(1.3)
$$
 where $\bar \nu$ is the {\it antineutrino.}

Despite the scientific authority of historical figures such as Pauli and Fermi, the conjecture on the
existence of the neutrinos as physical particles was never  accepted by the entire  scientific
community  because of: the impossibility for the neutrino to be directly detected in laboratory;
the neutrino inability to interact with matter in any appreciable way; and   the existence
of alternative theories that do not need the neutrino conjecture (see Refs. [5,6,7] and literature quoted
therein for earlier alternative theories).

By the middle of the 20-th century there was no clear experimental evidence
acceptable by the scientific community at large confirming the neutrino conjecture beyond
doubt, except for experimental claims in 1959 that are known today to be basically flawed on various grounds,
as we shall see in the next sections.

In the last part of the 20-th century, there was the advent of the so-called {\it unitary SU(3) theories}
that attempted  the Mendeleev-type classification of all strongly interacting particles
(known as {\it hadrons}) into families. Jointly said theories intended to characterize the structure of any
member of a given hadronic family  as a quantum bound state of  hypothetical particles known as {\it quarks}.
These studies were subsequently extended to the so-called SU(3) color and flavor theories and are more
recently known as the {\it standard model} of all elementary particles (see, e.g., Ref. [8] and vast
literature therein).

The validity of the classification of hadrons into families via unitary symmetries was established by clear
experimental confirmation of numerous predictions of new physical particles existing in our spacetime along
lines much similar to the historical predictions of new atoms achieved
by the Mendeleev classification, which prediction were all experimentally verified.

Nevertheless, the conjecture that quarks are physical particles
in our spacetime has never been widely accepted by the scientific community due to a plethora of
conceptual and technical inconsistencies (see refs.s [9,10]) some of which are outlined in the next section.

More recently, the controversies have been multiplied by the joining of neutrino and quark conjectures
because such a joining has required additional unverified conjectures. The standard model does not
predict the main features of the hypothetical neutrinos. This occurrence  is considered positive by
scientists accepting only the final classification character of the standard model, while the same occurrence
is considered a major drawback of the standard model by scientists accepting the conjecture that neutrinos
and quarks are physical particles.

The marriage of neutrino and quark conjectures within the standard model has requested the
multiplication of neutrinos, from the  neutrino and antineutrino conjectures of the early studies, into six
different hypothetical particles, the so called {\it electron, muon and tau neutrinos and their
antiparticles.} In the absence of these particles  the standard
model would lose the capability of providing both the classification and structure of particles.

In turn, the multiplication of the neutrino conjectures has requested the additional conjecture that the
electron, muon and tau neutrinos  have masses, plus the additional conjecture that they have
different masses, as necessary to salvage the structural features of the standard model. Still in turn,
the lack of resolution of the preceding conjectures has requested the
additional conjecture that {\it neutrinos  oscillate,} namely, that "they change flavor" (transform among
themselves back and forth).

In addition to this rather incredible litany of sequential conjectures, each conjecture being voiced in
support of a preceding  unverified conjecture, by far the biggest controversies have
occurred in regard to experimental claims of neutrino detection voiced by large collaborations.

To begin, both neutrinos and quarks cannot be directly detected as physical particles in our spacetime.
Consequently, all claims on their existence are indirect, that is, based on the detection of actual
physical particles predicted by the indicated theories. This occurrence is, per se, controversial. For
instance, controversies are still raging following  announcements by various laboratories to have "discovered"
one or another quark, while in reality the laboratories discovered physical particles predicted by a
Mendeleev-type classification of particles, the same classification being admitted by theories that
require no quarks at all, as well shall indicate later on.

In the 1980s, a large laboratory was built deep into the Gran Sasso mountain in Italy to detect neutrinos
coming from the opposite side of Earth since the mountain was used as a shield against cosmic rays.
Following the investment of large public funds, the Gran Sasso Laboratory released no evidence of clear
detection of neutrino originated events following five years of continuous tests.


\begin{figure}
\begin{center}
\epsfxsize=6cm
\parbox{\epsfxsize}{\epsffile{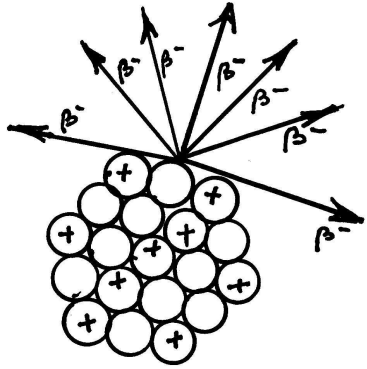}}
\caption{\small \it DISMISSAL OF THE NEUTRINO CONJECTURE DUE TO LACK OF ENERGY. Measurements done in the
first half of the 20-th century have established that the energy of the electron in nuclear beta decays has
a bell-shaped curve with the maximal value of $0.782 MeV$. the "missing energy" has been historically
attributed to be carried out by the hypothetical neutrino. However, as indicated in Section 2 (and treated
technically in the quoted literature), when the energy in the beta decay is computed with the inclusion of
the Coulomb interactions between the expelled (negatively charged) electron and the (positively charged)
nucleus at different directions of the expulsion, the nucleus acquires the "missing energy," without any
energy left for the hypothetical neutrino.}
\end{center}

\label{Fig1}
\end{figure}


Rather than passing to scientific caution in the use of public funds, the failure of the
 Gran Sasso experiments to produce any neutrino evidence stimulated new massive efforts by large
collaborations involving hundred of experimentalists from various countries. The increase in
experimental research was evidently due to the scientific stakes, because, as well known by experts,
the lack of verification of the neutrino conjectures would imply the identification of clear limits of
validity of quantum mechanics and special relativity.

These more recent experiments resulted in claims that, on strict scientific grounds, should be considered
"experimental beliefs" by serious scholars for numerous reasons, such as:

1) The
predictions are based on a litany of sequential conjectures none of which is experimentally established on
clear ground;

2) The theory contains a plethora of unrestricted parameters that can essentially fit any
pre-set data;

3) The "experimental results" are based on extremely few events out of hundreds of millions
of events over years of tests, thus being basically insufficient in number for any serious scientific
claim;

4) In various cases the "neutrino detectors" include radioactive isotopes that can themselves
account for the selected events;

5) The interpretation of the experimental
data via neutrino and quark conjectures is not unique, since there exist nowadays other theories
representing exactly the same events without neutrino and quark conjectures
(including a basically new
scattering theory of nonlocal type indicated later on).

Neutrino and quark conjectures have requested to date the expenditure of over one billion dollars of public
funds in theoretical and experimental research with the result of increasing the
controversies rather than resolving any of them.

Therefore, it is time for a moment of reflection: scientific ethics and accountability require that
serious scholars in the field  exercise caution prior to venturing claims of
actual physical existence of so controversial and directly unverifiable conjectures.

In a language as accessible to a broad audience as possible, and by delegating technical issues to quoted
references, in this note we
collect, apparently for the first time, the main arguments against the existence of neutrinos and quarks
as physical particles,  outline alternative theories without neutrino and quark conjectures, and point out
their negative environmental implication.

The predictable conclusion of this study is that theoretical and experimental research on neutrino and quark
conjectures should indeed continued. However, theoretical and
experimental research on theories without neutrino and quark conjectures and their new clean energies
should be equally supported to prevent a clear suppression of scientific democracy on fundamental issues,
evident problems of scientific accountability, and a potentially severe judgment by posterity.

\vskip0.50cm

\noindent {\bf 2. Catastrophic inconsistencies of neutrino and quarks conjectures.}

\noindent In regard to the neutrino conjecture, it is important to disprove it first as originally conceived,
and then disprove the flavored extension of the conjecture as requested by quark conjectures. The disproof of
the original conjecture of the neutrino can be done with the argument below and those of the following
sections. The disproof of the conjecture of flavored neutrinos is an inevitable consequence of the
catastrophic inconsistencies of the conjecture that quarks are physical particles.

As reported in nuclear physics textbooks, the energy experimentally measured as being carried by the electron
in beta decays is a bell-shaped curve with a maximum value of $0.782 MeV$, that is the difference in value
between the mass of the neutron and that of the resulting proton in decay (1.3).
The "missing energy" has been attributed throughout
the 20-th century to be carried out  by the hypothetical neutrino.

It is easy to see that these calculations were initially done to adapt reality to the neutrino conjecture
and then adopted by others without their critical examination. The electron in beta decays is
negatively charged, while the nucleus is positively charged. Consequently, {\it the electron in beta decays
experiences a Coulomb attraction from the original nucleus.}

Moreover, such an attraction is clearly
dependent on the angle of emission of the electron by a decaying peripheral neutron. The maximal value of the
energy occurs for radial emissions of the electron, the minimal value occurs
for tangential  emissions, and the intermediate value occur for intermediate directions of emissions,
resulting in the experimentally detected bell-shaped curve.


\begin{figure}
\begin{center}
\epsfxsize=7cm
\parbox{\epsfxsize}{\epsffile{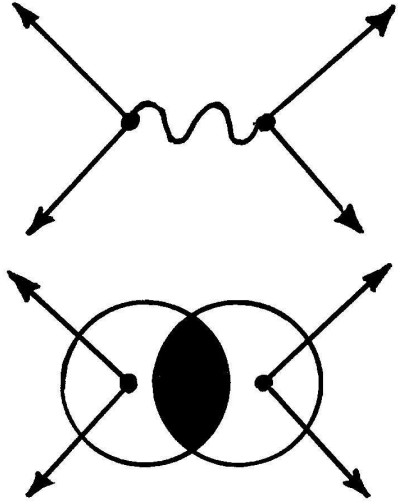}}
\caption{\small \it DISMISSAL OF THE NEUTRINO CONJECTURE DUE TO THE INAPPLICABILITY OF THE
QUANTUM SCATTERING THEORY. As indicated in Section 3 (and treated technically in the quoted literature),
the scattering theory used in all experimental claims on neutrino conjectures is structurally
insufficient for final scientific claims because it is  based on the abstraction of particles as massive
points (top view) that, as such, cannot have collisions. In reality, all hadrons are extended hyperdense
spheroids of radius $1 F = 10^{-13} cm$, and even though electrons have a point-like charge, they do not have
a "point-like wavepacket" as necessary for a serious validity of the quantum scattering theory. When
extended particles collide as in the physical reality, we have a volume of mutual penetration of their charge
distributions and/or wavepackets (lower view) that causes interactions of nonlinear, nonlocal, nonpotential
and nonunitary type, namely, interactions dramatically beyond any dream of treatment via quantum mechanics.
A covering nonlocal scattering theory has been built as part of the new  hadronic mechanics  outlined in
Section 4. It is easy to see that the elaboration of experimental data via a more realistic nonlocal
scattering theory produces results different than those claimed for neutrino  conjectures, thus
confirming the need for scientific caution prior to final claims that they are physical
particles.}
\end{center}
\end{figure}


When the calculations are done without prejudicial interests in existing doctrines, it is easy to see that
the "missing energy" in beta decays is entirely absorbed by the nucleus via its Coulomb interaction with
the emitted electron. Consequently, in beta decays there is no energy at all available for the  neutrino
conjecture, by reaching in this way a final disproof of the conjecture itself.

Supporters of the neutrino conjecture are expected to voice various
counter-arguments on the lack of experimental evidence for the nucleus to absorb said "missing energy."
These supporters are suggested to exercise scientific caution and first study the new structure models of the
neutron without the neutrino conjecture, as well as the resulting new structure models of nuclei before
venturing politically motivated views.

In regard to quark conjectures, the author has
repeatedly stated in his writing that {\it the unitary, Mendeleev-type,  SU(3)-color classification of hadron
into families has a final character.} All doubts herein considered solely refer to the joint use of the same
classification models as providing the structure of each individual element of a given hadronic family.

Consequently, all structure models considered, including those without neutrino and
quark conjectures, must achieve full compatibility with said unitary models of classification, in
essentially the same way according to which quantum structures of atoms achieved full
compatibility with their Mendeleev classification.

Far from being alone, this author has repeatedly expressed the view that {\it quarks cannot be physical
constituents of hadrons existing in our spacetime} for numerous independent reasons [9,10].

On historical grounds, the classification of nuclei, atoms and molecules required {\it two different
models,} one for the classification and a separate one for the structure of the individual elements of a
given family. Quark theories depart from this historical teaching because their conception of representing
with one single theory both the classification and the structure of hadrons.

As an example, the idea that the Mendeleev classification of atoms could jointly provide the structure of
each individual atom of a given valence family is outside the boundary of science. The Mendeleev
classification was achieved via {\it classical theories,} while the  understanding of the atomic
structure required   the construction of {\it a new theory,} quantum mechanics.

Independently from the above dichotomy classification vs structure, it is well known by technicians, but
rarely admitted, that {\it quarks are purely mathematical quantities, being purely mathematical
representations of a purely mathematical unitary symmetry defined on a purely mathematical
complex-valued unitary  space without any possibility, whether direct or implied, of being defined in
our spacetime (technically prohibited by the O'Rafearthaigh theorem).}

It follows that the conjecture that quarks are physical particles is afflicted by a plethora of major
problematic aspects today known to experts as {\it catastrophic inconsistencies of quark conjectures,} such
as:

1) No particle possessing the peculiar features of quark conjectures (fraction charge, etc.) has ever been
detected throughout the 20-th centuries in  any high energy physical laboratory around the world.
Consequently, a main consistency requirement of quark conjectures is that quarks cannot be produced free
even under the extremely high energies achived by current particle laboratories and, consequently, they must
be "permanently confined" in the interior of hadrons even. However, it is  well known to experts that,
despite half a century of attempts, {\it no truly convincing "confinement of quarks" inside protons and
neutrons has been achieved to date,} nor it can be expected on serious scientific grounds  by assuming (as
it is the case of quark theories) that quantum mechanics is identically valid inside and outside hadrons.
This is due to a pillar of quantum mechanics, Heisenberg's uncertainty principle, according to which, given
any manipulated theory appearing to show confinement for a given quark, a graduate student in physics can
prove the existence of a finite probability for the same quark to be outside the hadron {\it at low
energies}, thus being free, while the probability for the production of free quarks at very high energy can
only be defined as being embarassing. These fact establish beyond "credible" doubt that quark conkjectures
are  in catastrophic disagreement with physical reality. Consequently, the conjecture
that quarks are physical particles is afflicted by catastrophic inconsistencies in its very conception since
all quark theories predict the production of free quarks under sufficiently energetic collisions, while no
quark has ever been detected free and no true confinement is possible on serious scientific grounds accepted
by the scientific community at large [9].

2) It is equally well known by experts that {\it quarks cannot experience gravity} because quarks
cannot be defined in our spacetime, while  gravity can only be formulated in our
spacetime and does not exist in mathematical complex-unitary spaces. Consequently, if protons and neutrons
were indeed formed of quarks, we would have the catastrophic inconsistency that all quark believers should
float in space due to the absence of gravity [10].

3) It is also well known by experts that {\it "quark masses" cannot possess any inertia} since they are
purely mathematical parameters that cannot be defined in our spacetime. A  condition for any
mass to be physical, that is, to have inertia, is that it has to be the eigenvalue of a Casimir invariant
of the Poincar\'e symmetry, while quarks cannot be defined via said symmetry because of their
hypothetical fractional charges and other esoteric assumptions. This aspect alone implies numerous
catastrophic inconsistencies, such as the impossibility to have the energy equivalence $E = mc^2$ for any
particle composed of quarks, against vast experimental evidence to the contrary [10].

4) Even assuming that, because of some twist of scientific manipulation, the above inconsistencies are
resolved, it is known by experts that quark theories have failed to achieve a representation of all
characteristics of protons and neutron, with catastrophic inconsistencies in the representation of spin,
magnetic moment, mean lives, charge radii and other basic features.

5) It is also known by experts that the application of quark conjectures to the structure of nuclei has
multiplied the controversies, while resolving none of them. As an example, the assumption
that quarks are the constituents of the protons and the neutrons composing nuclei has failed to achieved a
representation of the main characteristics of the simplest possible nucleus, the deuteron. In fact, quark
conjectures are afflicted by the catastrophic inconsistencies of being unable to represent the spin 1 of
the deuteron (since they predict spin zero in the ground state), they are
unable to represent the anomalous magnetic moment of the deuteron, they are unable to represent the deuteron
stability, they are unable to represent the charge radius of the deuteron, and when passing to larger nuclei,
such as the zirconium, the catastrophic inconsistencies of quark conjectures can only be defined as being
embarrassing.

In summary, while the final character of the SU(3)-color classification of hadrons into families has
reached a value beyond scientific doubt, the conjecture that quarks are the actual physical constituents of
hadrons existing in our spacetime is afflicted by so many and so  problematic aspects to raise
serious issues of scientific ethics and accountability, particularly in view of the ongoing
large expenditures of public funds in the field.

It then follows that any additional conjecture based on the quark conjecture, such as that of the electron,
muon and tau neutrinos and related additional conjecture of their oscillations, are so clearly flowed not to
warrant detailed dismissals because unresolved conjectures cannot be credibly
bypassed with additional conjectures, while the rather frequent use of academic power and credibility to
support unverifiable conjectures  can only multiply the controversies.

Above all, the most serious problems of scientific ethics and accountability emerge when the  belief of
neutrino and quark conjecture emerge as having serious negative impact on major environmental issues,
as illustrated in Section 6.
\vskip0.50cm

\noindent {\bf 3. The inapplicability of quantum mechanics inside hadrons.}

\noindent The proton and the electron are the only massive stable particles clearly identified until now.
Consequently,  the idea that, as necessary for the standard model, the proton and the electron must
"disappear" in synthesis (1.1) and be transformed into hypothetical undetectable quarks is repugnant to
reason.

The most
logical expectation is that, being stable particles, the proton and the electron persist in synthesis (1.1)
according to Rutherford's original conception [1]. This is  why, despite the successes of
weak interactions, studies along Rutherford's original conception
have continued since the times of Ref. [1].

The main obstacle against Rutherford's conception of the neutron is that it is prohibited by quantum
mechanics, as pointed out by Pauli, Fermi, Schroedinger and other founders of quantum mechanics, who voiced
the following serious objections:

1) Quantum mechanics cannot represent the spin of the neutron under Rutherford's conception because
the total angular momentum of the ground state of a two particles with spin 1/2, such as the proton and
the electron, must be 0, while the neutron has spin 1/2.

2) The representation of synthesis (1.1) via
quantum mechanics is impossible because it would require a "positive binding energy," in violation of basic
quantum laws requiring that all binding energies must be negative, as proved in nuclear physics. This is
due to the fact that the sum of the mass of the proton and of the electron,
$$
m_p + m_e  = 938.272 Mev +  0.511 MeV = 938.783 MeV,
\eqno(3.1)
$$
is {\it smaller} than the mass of the neutron, $m_n = 939.565 MeV$, with "positive mass defect"
$$
m_n - (m_p + m_e) = 939.565 - (938.272 +  0.511) MeV = 0.782 MeV.
\eqno(3.2)
$$
Consequently, quantum mechanics
would require a positive binding energy bigger than $0.782 MeV$, under which  all quantum equations
become inconsistent (technically, the indicial equations of Schroedinger's equation becomes inconsistent
and prevents the representation of the total energy with real numbers).

3) Via the use of the magnetic moment of the proton $\mu_p = 2.792 \mu_N$ and of the electron $\mu_e = 1.001
\mu_B$, it is impossible to reach the magnetic moment of the neutron $\mu_n = - 1.913 \mu_N$.

4) When the neutron is interpreted as a bound state of one proton and one electron, it is impossible to
reach the neutron meanlife $\tau_n = 918 \;\;sec$ that is quite large for particle standards, since quantum
mechanics would predict the expulsion of the electron in nanoseconds.

5) There is no possibility for quantum mechanics to represent the neutron radius of about $1 F = 10^{-13}
cm$ since the smallest predicted radius is that of the hydrogen atom of $10^{-8} cm$, namely 5,000 times
{\it bigger} than the neutron radius.

The above objections were indeed valid at the time of their formulations in the
1940s. However, due to the
 knowledge gained on the structure of the proton and neutron since that time, the above objections are
nowadays no longer valid for numerous reasons.


\begin{figure}
\begin{center}
\epsfxsize=7cm
\parbox{\epsfxsize}{\epsffile{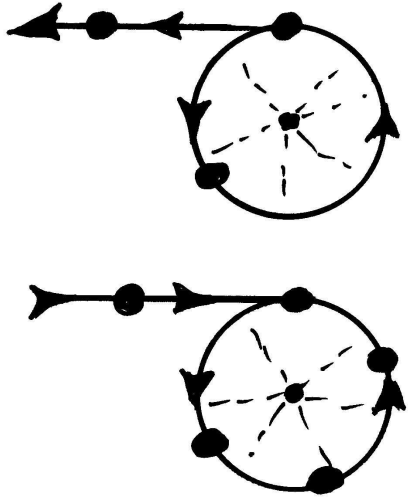}}
%
\caption{\small \it DISMISSAL OF THE NEUTRINO CONJECTURE DUE TO THE LACK OF ANGULAR MOMENTUM CONSERVATION IN
SCATTERINGS OF EXTENDED PARTICLES. As it is well known, the neutrino conjecture was formulated on the basis
of the quantum mechanical conservation of angular momentum. All experimental claims on neutrino
conjectures are based on the Poincar\'e invariant scattering theory that, in turn, is based on the angular
momentum conservation law. However, it is equally well known that said conservation only occurs for
planetary-atomic systems, namely for Keplerian systems of point particles in stable orbits without
collisions. In the scattering of extended particles only the total energy is conserved since the angular
momentum can transform itself in linear momentum and vice versa, as illustrated in this picture with the
sling-shot, or as anybody has eyewitnessed in the scattering of billiard balls. The lack of conservation of
angular momentum eliminates the very foundations for the original proposal of the neutrino conjecture. }
\label{Fig3}
\end{center}
\end{figure}


Quantum mechanics was conceived and constructed for the representation of the trajectories of electrons
moving in vacuum in atomic orbits, in which field the theory received historical verifications.
The same mechanics cannot possibly be exact for the description of the dramatically different physical
conditions of the same electron moving within the hyperdense medium inside a proton. Such an
assumption literally implies the belief in the perpetual motion within a physical medium since it implies
that an electron must orbit in the core of a star with a conserved angular momentum, as requested by the
quantum axiom of the rotational symmetry and angular momentum conservation law.

In addition, there exist today a rather large body of evidence according to which quantum mechanics,
while being exactly valid under the
conditions of its conception (point-like particles moving in vacuum), is {\it inapplicable} wiuthin hadrons,
with the understanding that claims of "violation of quantum mechanics inside hadrons" would not be
scientific, first of all, because the theory was not constructed for the field considered and, second,
because all theories can at best approximate nature.

For comprehensive studies of this issue we refer the reader to monographs [16] and vast literature quoted
therein. In this note it is sufficient to recall that quantum mechanics cannot be exact inside hadrons
because its fundamental Galileo and Poincar\'e symmetries are broken since they were conceived for motion in
vacuum and certainly not within hyperdense media.

As it is well known, the validity of said basic symmetries requires the validity of the {\it celebrated ten
conservation laws of the total energy, linear momentum, angular momentum and uniform motion of the center
of mass} that can be classically expressed with the unified law
$$
\frac{dX_i(t, r, p)}{dt} = \frac{\partial X_i}{\partial b^\mu}\times \frac{db^\mu}{dt} +
\frac {\partial X_i}{\partial t} = 0,
\eqno(3.3)
$$
where
$$
X_1 = E_{tot} = H = T + V,
\eqno(3.4a)
$$
$$
(X_2, X_3, X_4) = {\bf P}_{Tot} = \Sigma_a {\bf p}_a,
\eqno(3.4b)
$$
$$
(X_5, X_6, X_7) = {\bf J}_{tot} = \Sigma_a {\bf r_a} \wedge {\bf p}_a,
\eqno(3.4c)
$$
$$
(X_8, X_9, X_{10}) = {\bf G}_{Tot} = \Sigma_a (m_a\times {\bf r}_a -t\times {\bf p}_a),
\eqno(3.4d)
$$
$$
 i = 1, 2, 3, ..., 10; \; \;k = 1, 2, 3; \; \; \; a = 1, 2, 3, ..., N,
$$
with corresponding quantum expressions here ignored for simplicity.

It bis easy to see that, while all the above  conservation laws are indeed verified for planetary-atomis
structures, {\it the sole conservation law valid for physical systems in general is the conservation
of the energy, due to the known lack of conservation of the linear and angular momentum in
actual collisions.}

There is no need for high energy laboratories to see this occurrence since it is sufficient to drop a mass
on a table. As anyone can observe, in this collision the linear momentum {\it cannot}
be conserved. The only conserved quantity  is the kinetic energy that, in this
case, is transformed into heat energy.

Similarly, Figure 3 illustrates the exchange of linear and angular momenta, but always under the
conservation of the energy, since we merely have in this case the transformation of kinetic energy into
rotational energy and vice versa.

Note that {\it the above physical collisions among extended objects are sufficient to
invalidate the experimental claims that neutrinos and quarks are physical particles since said claims all
result from collision of extended particles.} Alternatively, the approximate character of quantum
mechanics in hadron physics can be seen from the fact that {\it the quantum scattering theory cannot
be exactly valid for particle collisions since it must represent the particles as massive points, and
points cannot collide because they are dimensionless.} Under these extreme abstractions, how can
"experimental beliefs" be turned into physical reality?

Unreassuringly, the above arguments are only the beginning of the litany of strong evidence preventing
quantum mechanics from being exact for the hadronic structure.
It is equally well known by expert, but not sufficiently spoken, that the Galileo and Poincar\'e symmetries
can be exactly valid only under the conditions that:

I) All constituents can be effectively approximated as
massive points (a condition necessary from the very mathematical structure of the symmetries, beginning from
the underlying topology);

II) The constituents move in stable and quantized orbits without collisions (otherwise the structural
axioms would not apply); and

III) The system considered must exhibit a {\it Keplerian center,} namely, the heaviest element is located
in one of the two foci (otherwise the Keplerian structure cannot occur).

It is easy to see that {\it none of the above central conditions for the validity of the Galileo and
Poincar\'e symmetries are valid inside hadrons, with resulting lack of exact validity of the Galileo and
Poincar\'e symmetries and the consequential lack of exact character of quantum mechanics.}  In fact:

I*) The constituents of hadrons cannot be effectively approximated as being point-like because that would
require {\it hadronic constituents to have "pointlike wavepackets,"}  something without any
scientific sense. All particles, beginning with the electrons have
non-null wavepackets that are rather large for particle standards because of approximately the size of all
hadrons (1 F). Consequently,  hadronic constituents are in a state of total mutual penetration of their
wavepackets, resulting in {\it nonlocal interactions} (that is, interactions extended over a volume) that are
beyond any possibility of quantum description since quantum theories can only represent events occurring
among isolated  points.


\begin{figure}
\begin{center}
\epsfxsize=12cm
\parbox{\epsfxsize}{\epsffile{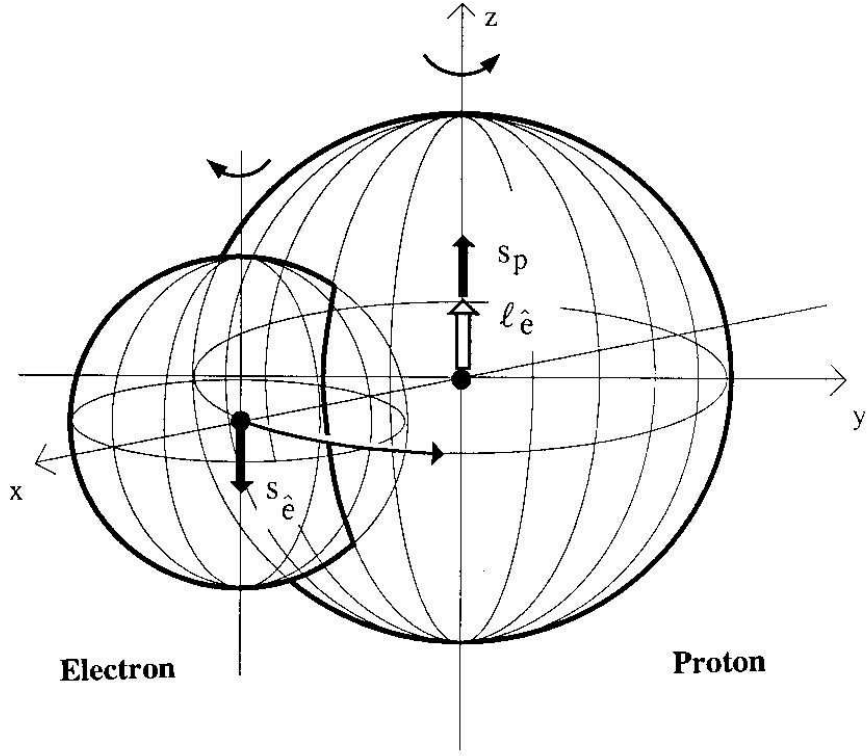}}
\caption{\small \it ABSENCE OF THE NEUTRINO CONJECTURE IN RUTHERFORD'S SYNTHESIS OF THE NEUTRON ACCORDING TO
HADRONIC MECHANICS. At the initiation of Rutherford's compression of the hydrogen atom in the core of a
star, when the electron penetrates within the hyperdense medium inside the proton it is constrained for
stability to couple with antiparallel spins, while it is equally constrained to have an angular momentum that
coincides with the spin of the proton, as illustrated in the figure. In this process the electron is mutated
into the isoelectron, that is, a particle characterized by the covering Poincar\'e-Santilli isosymmetry.
Consequently, at the completion of the compression, the isoelectron has null total angular momentum and the
spin of the neutron coincides with that of the proton without any need for the neutrino conjecture. It
should be indicated that fractional values of angular momenta are anathema for quantum mechanics because
they break its unitary structure. However, the same values are normal for the covering hadronic mechanics
since the latter theory has been build for the invariant treatment of nonunitary structures.}
\label{Fig4}
\end{center}
\end{figure}


II*) The orbits of hadronic constituents are indeed stable but they cannot be  quantized as for the
orbit of atomic electrons, again, because motion occurs within a hyperdense medium. This is established by
the fact that any quantum jump from one orbit to another would mandate the constituent to exit the hadron,
since the minimal spacing in quantum orbits is greater than the size of hadrons. In different terms,
the admission of quantized orbits for quark constituents of the proton mandates that the proton is unstable
and the work is free because, when excited, said quark must jump to a new orbit necessarily outside the
proton itself, in catastrophic disagreement with experimental reality.

III*) The Galileo and Poincar\'e symmetries {\it cannot} be exactly valid for the hadronic structure because
{\it hadrons do not possess a Keplerian centers} as it is the case for atoms. Consequently, hadrons
cannot constitute Keplerian systems, and quantum mechanics cannot be exact. According to
incontrovertible data from deep inelastic scattering, hadrons are composed of a somewhat homogeneous and
isotropic hyperdense medium in which the search for the remnants of an atomic structure has no scientific
sense.

Note that the absence of an atomic structure inside the proton is sufficient, per se, to invalidate all
arguments against Rutherford's conception of the neutron, since all these arguments
are based on said atomic structure.

In the final analysis it has been established by scientific history that the validity of any given theory
within given conditions is set by the results. Quantum mechanics has represented {\it all} features of the
hydrogen atom in a majestic way and, therefore, the theory is exactly valid within the indicated
conditions. By contrast, when extended to the structure of particles, quantum mechanics has only produced
an interlocked chain of individually unverifiable conjectures on neutrinos and quarks, besides failing to
achieve final results in various other branches of sciences, such as in nuclear physics, chemistry and
astrophysics [16].

Rather than continuing with additional unverifiable conjecture, it is time to re-examine the validity of
quantum mechanics for the structure of hadrons. After all these dramatic controversies protracted for such
a long period of time, there comes a point in time in which the serious conduction of serious science
requires a re-examination of the foundations.

\vskip0.50cm

\noindent {\bf 4. The structure model of hadrons with massive physical constituents.}

\noindent The view advocated by the author is that a deeper understanding of the structure of hadrons
requires a generalization of the quantum  theory valid for their classification, because  a theory
constructed for the description of electrons orbiting in vacuum in atomic structures, cannot be credibly
claimed to be necessarily valid for the description of the same electrons when moving within the hyperdense
media inside hadrons.

In view of all the insufficiencies and controversies in various branches of sciences, R. M. Santilli [20]
proposed in 1978 the construction of a generalization/covering of quantum mechanics under the name of  {\it
hadronic mechanics} precisely to represent the dynamics within hyperdense media inside hadrons, under
the conditions of achieving full compatibility with quantum theories of classification, as well as admitting
quantum mechanics uniquely and unambiguously at the limit when all resistive forces caused by motion within
hyperdense media cease to exist and motion returns to be in vacuum.

 Following proposal [20], hadronic mechanics was
studied by numerous mathematicians, theoretician and experimentalists (see monographs [11-19,33-39] and vast
literature quoted therein). After decades of studies, mathematical maturity in the formulation of hadronic
mechanics was reached only in 1996 [21], after which it was easy to achieve physical maturity [22,23].

As we shall see, hadronic mechanics has permitted the exact, numerical and invariant representation of
{\it all} characteristics of hadrons without any neutrino and quark conjecture and via the use of ordinary
massive particles as physical constituents in our spacetime, generally those produced free in the
spontaneous decays with the lowest mode.

In particular, as clearly stated in the original proposal [20],  hadronic mechanics was proposed for the
specific intent of achieving a consistent quantitative representation of Rutherford's synthesis (1.1) by
resolving all the objections outlined in Section 3. This objective was set  in view of the important
environmental implications outlined in the next section.

The main idea of proposal [20] is essentially the following. One of the axioms of quantum mechanics is that
its time evolution must characterize a {\it unitary transform} in a Hilbert space $\cal {H}$ over the field
of complex numbers $\cal{C}$,
$$
U\times U^\dag = U^\dag\times U =  I.
\eqno(4.1)
$$

A necessary condition to achieve a true generalization of quantum mechanics is then that of exiting from its
class of all possible equivalent formulations. Consequently, a central axiom of
hadronic mechanics is that its time evolution must characterize a {\it nonunitary transform} when
expressed on $\cal {H}$ over $\cal {C}$,
$$
U\times U^\dag \not =  I.
\eqno(4.2)
$$

However, it was known in 1978 that nonunitary theories do not have consistent physical predictions when
treated via the mathematics of quantum mechanics, such as Hilbert spaces $\cal {H}$ and fields $\cal
{C}$. This mandated the construction of a {\it basically new
mathematics} specifically conceived for physically consistent treatment of nonunitary theories.

The solution proposed in the original memoirs [20] is the generalization (called {\it lifting})
of the trivial unit $I = +1$ of quantum mechanics into a positive-definite, but most general possible
integro-differential operator $\hat I(t, r, p, \psi, ...)$, today called {\it Santilli's isounit,} that is
assumed to coincide with nonunitary transform (4.2),
$$
U\times U^\dag  =  \hat I(t, r, p, \psi, ...) = 1 / T(t, r, p, \psi, ...) > 0
\eqno(4.3)
$$
In turn, the lifting $I\rightarrow \hat I$ required a generalization of the conventional associative
product $A \times B$ between generic quantities A, B (numbers, matrices, etc.)  into the form
$$
A\hat {\times} B = A\times \hat T\times B,
\eqno(4.4)
$$
under which $\hat I$ is indeed the correct left and right unit
$$
\hat I\hat \times \hat A = \hat I\times \hat T\times A = A\hat \times \hat I = A,
\eqno(4.5)
$$
for all elements A of the set considered.

As a pre-requisite for practical applications of hadronic mechanics, the lifting of the unit
and of the product required the construction of a compatible  generalization of the totality of the
mathematics of quantum mechanics into a covering formulation today known as {\it Santilli's isomathematice,}
that includes generalized numbers, fields, metric spaces, topologies, functional analysis, algebras,
geometries, symmetries, etc.  This explains the decades of  work that were necessary to achieve maturity of
applications of hadronic mechanicsd.,

On physical grounds,  the generalized unit $\hat I(t, r, p, \psi, ...)$ is used for the invariant
representation of {\it contact forces among extended constituents} that are completely absent in
quantum mechanics.

Santilli [20] selected the unit for the representation of contact interactions among extended particles
because, on one side, the latter are not potential, thus being outside the quantum Hamiltonian, and, on the
other side, because the unit is the {\it only} alternatieve permitting an invariant representation.
In fact, whether conventional or generalized, the unit is the basic invariant of any theory.

Via the use of hadronic mechanics, the original proposal [20] achieved already in 1978 a new structure model
of mesons with actual massive physical  constituents according to the models
$$
\pi^o = (\hat e^+, \hat e^-)_{HM},
\eqno(4.6a)
$$
$$
\pi^\pm = (\hat e^+ \hat e^\pm, \hat e^-)_{HM},
\eqno(4.6b)
$$
$$
 K^o = (\hat \pi^+,
\hat \pi^-)_{HM}, etc.,
\eqno(4.6c)
$$
where $e, \pi, K, etc.$ represent conventional particles as detected in laboratory and $\hat
e, \hat \pi, \hat K, etc.$ represent their {\it mutation,} that is, the alternation of their characteristics
when in deep mutual penetration of their wavepackets and charge distributions (a feature tecxhnically
treated via representations of the Galileo-Santilli [13] and Poincar\'e-Santilli [14] isosymmetries).

Hadronic structure models (4.6) achieved an exact, numerical and invariant representation of {\it all}
characteristics of  mesons, including characteristics whose representation has been
impossible for quark conjectures after some three decades of failures, such as the representation of: 1) the
charge radius of the particles; 2) Their meanlife, and 3) the reason the massive constituents are
emitted in the spontaneous decays with the lowest mode.

Hadronic mechanics can now be constructed quite simply by applying a nonunitary transform to {\it all}
quantities and their operations of quantum mechanics, including number, metric spaces, algebras, geometries,
topologies, etc. Recall that the quantum model underlying the hadronic structure of the $\pi^o$, Eq. (4.6.a),
is the positronium (a bound state at large mutual distances of an electron and a positron). Therefore,
model (4.6a) can be easily achieved via the lifting of the qwuantum Scroedinger's and Heisenber's equations
$$
U\times U^\dag = \hat I = 1 / \hat T \not = 1,
\eqno(4.7a)
$$
$$
i\times U\times {dA\over dt}\times U^\dag = i\times {\hat d\hat A\over \hat d\hat t} = U\times [A, H]\times
U^\dag =
$$
$$
= \hat A\times \hat T\times \hat H - \hat H\times \hat T \times \hat A =  \hat A\hat \times \hat H - \hat
H\hat \times \hat A = [\hat A\hat {,} \hat H],
\eqno(4.7b)
$$
$$
U\times (H\times |\psi>) = (U\times H\times U^\dag)\times (U\times U^\dag)^{-1} \times (U\times |\psi>) =
$$
$$
= \hat H\times \hat T\times |\hat \psi> = \hat H\hat \times |\hat \psi> = U\times (E\times |\psi) = \hat
E'\hat \times |\hat \psi> = E'\times |\hat \psi>,
\eqno(4.7c)
$$
$$
\hat H = U\times H\times U^\dag, \hat A = U\times A\times U^\dag, |\hat \psi>  = U\times |\psi>,
\eqno(4.8b)
$$
where: the isotopic generalization (4,7b) of Heisenberg's equation is the fundamental equation
of hadronic mechanics proposed in the original memoirs [20] of 1978; Eq. (4.7b) is the isotopic
generalization of Schroedinger's equations achieved subsequently; H and
$|\psi>$ are  the Hamiltonmian and wavefunction, respectively, of the positronium; their nonunitary images
$\hat H$ and $|\hat \psi>$ are the corresponding expressions for the
$\pi^o$; $d/dt$ is the conventional differential; and $\hat d/\hat d\hat t$ is the isodifferential [21].

Rather remarkably, the representation was reached via the following simple isounit
$$
U\times U^\dag = \hat I = e^{k\times (\psi/\hat \psi)\times \int dr^3\times \psi^\dag(r)_{\uparrow}\times
\psi(r)_{\downarrow}},
\eqno(4.8)
$$
where k is a normalization constant and  $\int dr^3\times \psi^\dag(r)_{\uparrow}\times
\psi(r)_{\downarrow},$ represent the {\it nonlocal interactions} caused by the deep overlapping of the
wavepackets of the electron and positron  one inside the other,  a feature that is outside any possible
representation via quantum mechanics.

At the limit
$$
\int dr^3\times \psi^\dag(r)_{\uparrow}\times
\psi(r)_{\downarrow}, \rightarrow 0,
\eqno(4.9)
$$
namely, at the limit of large (atomic) distances and ignorable waveoverlapping,  hadronic model (4.6a)
recovers the quantum model of the positronium exactly, uniquely and unambiguously.

The use of quantum mechanics for models (4.6) yield the inconsistencies 2) to 5) of Section 3  but
inconsistency 1) required no treatment for  mesons because in this case there is no violation of the
quantum conservation of angular momentum, since the spin of the mesons can be achieved via the spin of their
massive constituents. Consequently, it was relatively easy to reach models (4.6) in the original proposal
[20] to build hadronic mechanics.

The representation of  Rutherford's synthesis of the neutron, Eq. (1.1), required considerable additional
studies on the isotopies of angular momentum and spin (see   papers [24,25]). After decades from the
original proposal [20], {\it  a nonrelativistic, exact, numerical and invariant representation of all
characteristics of the neutron as a bound state of a proton and an electron was finally achieved in papers
[26] of 1990, with the relativistic extension achieved in paper [27] of 11993 (see also Ref. [28]),}
according to the expression
$$
n = (\hat p^+, \hat e^-)_{HM},
\eqno(4.10)
$$
where $\hat p^+$ and $\hat e^-$ are the conventional proton and electron in a deformed (mutated) form caused
by their deep mutual penetration called {\it isoprotons and isoelectrons} (technically represented via
irreducible representation of the Poincar\'e-Santilli isosymmetry).

It should be noted that, since the proton is about 2,000 times heavier than the
electron, all calculations can be done with the simpler model
$$
n = (p^+, \hat e^-)_{HM},
\eqno(4.11)
$$
namely, the sole deformation of the electron caused by its Rutherford's compression inside the hyperdense
proton is sufficient to represent {\it all} features of the neutron.


\begin{figure}
\begin{center}
\epsfxsize=12cm
\parbox{\epsfxsize}{\epsffile{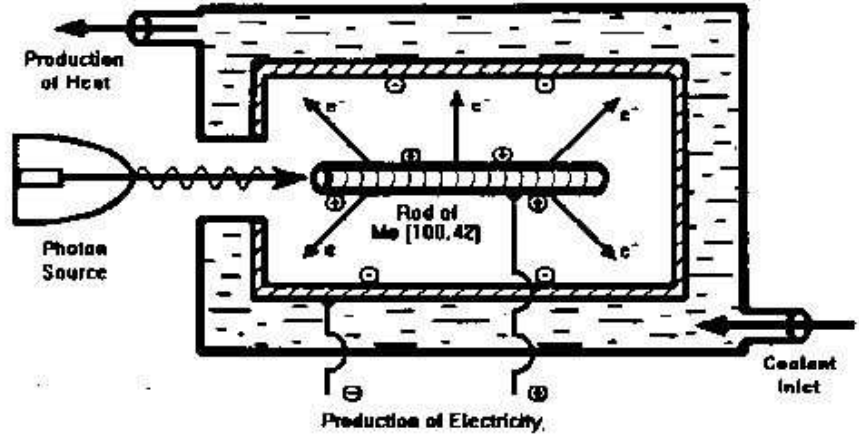}}
\caption{\small \it THE PREDICTED CLEAN AND INEXHAUSTIBLE ENERGY FROM THE STIMULATED DECAY OF RUTHERFORD'S
NEUTRON. The neutron is, by far, the largest and inextinguishable source of clean energy available to
mankind because it releases a high energetic electron with up to $0.782 MeV$ that can be easily trapped with
a thin metal shield, while the hypothetical neutrino, assuming that it exists, is harmless to humans and the
environment. If the neutron is composed of hypothetical quarks, there is no possibility whatever to tap
this clean energy, trivially, because the conjectural quarks cannot be produced free. The same holds if the
neutrino exist because the neutrino hypothesis was developed precisely to prevent that the electron is a
physical constituent of the neutron. On the contrary, if the neutron has Rutherford's structure, its energy
can be utilized, e.g., via its stimulated decay originating from the excitation of the electron. The
above picture is a reproduction of the original figure of Ref. [29] to illustrate that the produced energy is
twofold, electric energy in the form of a continuous current (called hadronic battery) caused by the
difference of potential between the electron shield and the original metallic fuel, as well as heat acquired
by said shield. Hadronic mechanics has predicted the exact value of the resonating frequency for the
stimulated decay of the neutron, the restricted class of nuclear isotopes admitting such a stimulated decay
(hadronic fuel) and  other aspects needed for industrial; development [29]. The energy produced is predicted
to be a large multiple of the energy used since a high efficiency can be achieved with a small submultiple
of the resonating frequency. This energy, the first predicted to originate from industrial hadrons, rather
than nuclei, is clean because it releases no harmful radiations and leaves no harmful waste, as
illustrated in the text [see Refs. [16,26] for details).}
\label{Fig5}
\end{center}
\end{figure}


The exact, numerical and invariant representation of the rest energy, meanlife, charge radius,  anomalous
magnetic moments and parities were achieved as for models (4.6). The representation of the spin 1/2 of the
neutron turned out to be much simpler than expected, as outlined in Figure 4.

In particular, {\it the hadronic representation of the synthesis of the neutron does not require any
neutrino conjecture at all, exactly as originally conceived by Rutherford.} Once compressed inside the
proton, the electron is constrained for stability to have its spin antiparallel to that of the
proton and its orbital angular momentum  to coincide with the spin 1/2 of the proton,
resulting in the following exact representation (where the proton and the resulting electrons are assumed to
be at rest for simplicity)
$$
s_n^{spin} = s_p^{spin} + s_{\hat e}^{spin} + s_{\hat e}^{orb} = {1\over 2} - {1\over 2} + {1\over 2} =
{1\over 2},
\eqno(4.12)
$$
namely, {\it the total angular momentum of the isoelectron is null},
$$
s_{\hat e}^{tot} = s_{\hat e}^{spin} + s_{\hat e}^{orb} = 0,
\eqno(4.13)
$$
and {\it the spin of the neutron coincides with that of the proton.}

Remarkably, after taking into account all contributions, configurations (4.12) resulted to yield an exact,
numerical and invariant representation of the anomalous magnetic moment of the neutron, thus providing
final confirmation [26,27,28].

It should be noted that a fractional  angular momentum is pure anathema for quantum mechanics
(because it implies a departure from the nonunitary structure of the theory with a host of problems), while
the same value is perfectly normal for hadronic mechanics defined on an iso-Hilbert space $\hat {\cal{
H}}$ over an isofield $\hat {\cal { C}}$ (see Refs. [26-28] for technical aspects)

As a matter
of fact, the nonunitary character of fractional values of orbital angular momenta is a direct confirmation of
the need for a nonunitary theory. Hadronic mechanics is the {\it only} known nonunitary theory that has
reached invariance and axiomatic consistency, besides a large number of experimental verifications in
various fields [11-39].

As one can see, the spontaneous decay of the neutron into physical, actually observed particles
$$
n = (p^+, \hat e^-)_{HM} \rightarrow p^+ + e^-,
\eqno(4.14)
$$
is a mere tunnel effect of the massive constituents. Assuming that the neutron and the proton are {\it
isolated} and at rest in the spontaneous decay, the electron is emitted with $0.782 MeV$ energy. When the
decaying neutron is a member of a nuclear structure, the energy possessed by the electron is generally less
than $0.782 MeV$ and varies depending on the angle of emission, as indicated in Figure 1 and Section 2.

Again, hadronic mechanics permitted the exact, numerical and invariant representation of features whose
treatment with quark conjectures is impossible, such as the representation of: 1) The charge radius
$10^{-13} cm$ of the neutron, 2) Its mean life $\tau_n = 918 sec$, and the reason the massive constituents
are emitted in the spontaneous decay.

The extension of the model to all remaining unstable particles resulted to be elementary.
Finally,  full compatibility of the above hadronic structure models with the SU(3)-color Mendeleev-type
classification of hadrons resulted to be possible in a variety of ways, such as, via a multivalued isounit,
as expressed below for the meson octet
$$
\hat I = Diag. (\hat I_{\pi^o}, \hat I_{\pi^+}, \hat I_{\pi^-}, \hat I_{K^o_S}, ... ).
\eqno(4.15)
$$
(see Ref. [14,16,22] for brevity).

The conclusion beyond scientific or otherwise credible doubts is that {\it all
features of all unstable (thus composite) particles can be uniquely and unambiguously represented in an
exact, numerical and invariant way as hadronic bound states of massive constituents generally produced free
in the spontaneous decays with the lowest mode, including the representation of various features that
are impossible with quark conjectures, said hadronic structure models being fully compatible with the
unitary Mendeleev-type classification of hadrons.}

\vskip0.50cm

\noindent {\bf 5. The negative environmental impact of neutrino and quark conjectures.}

\noindent Molecular, atomic and nuclear structures have provided immense benefits to mankind because their
constituents can be produced free. Quark conjectures on the structure of hadrons have no practical value
whatever, not even remote, because, by comparison, quarks cannot be produced free.

On the contrary, the
structure model of hadrons based on physical constituents that can be produced free have predicted new clean
energies originating in the structure of individual hadrons, today known as {\it hadronic energies.}

As an illustration, {\it the neutron is the biggest reservoir of clean energy available to
mankind} because: 1) The neutron is naturally unstable; 2) When decaying, it releases a large amount of
energy ($0.782 MeV$) carried out by the emitted electron; and, most importantly, 3) Such energy is
clean because the electron can be trapped with a thin metal shield, the energy thus
being without dangerous radiations and  without
dangerous waste as occurring for the nuclear energy.

Moreover, the latter type of hadronic energy is  two-fold because, when the decay of the
neutron occurs in a conductor, the latter acquires a positive charge while the shield
trapping the electron acquires a negative charge, resulting in a new clean production of continuous current
first proposed in Ref. [29] and today known as {\it Santilli's hadronic battery.} The second
source of energy is thermal and it is given by the  heat acquired by the  shield trapping the emitted
electrons.


\begin{figure}
\begin{center}
\epsfxsize=9cm
\parbox{\epsfxsize}{\epsffile{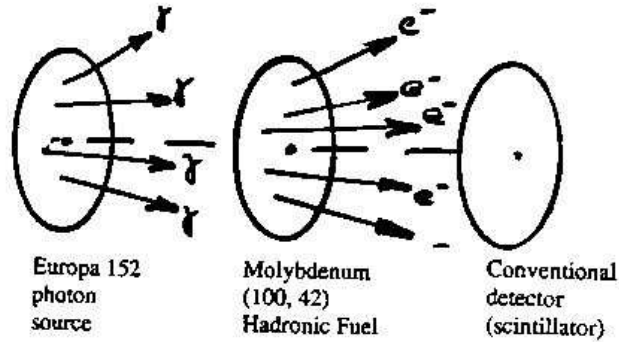}}
\caption{\small \it A SCHEMATIC VIEW OF TSAGAS EXPERIMENT ON THE STIMULATED DECAY OF THE NEUTRON. The
selected "hadronic fuel" is the MO(100,42) that, when hit by photons with the resonating frequency (5.1b) is
predicted to experience a stimulated decay into an unstable isotope that, in turn, decays spontaneous into a
final stable isotope with the total emission of two highly energetic electron, Eqs. (5.3),thus realizing the
conditions of Figure 5 with a large gain of energy (see Refs [16,26] for details).}
\label{Fig6}
\end{center}
\end{figure}


The main motivation of this paper is the indication that {\it the current widespread belief on neutrino
and quark conjectures prevents the orderly scientific study on how to utilize the
inexstinguishable clean energy inside the neutron, thus raising serious problems of scientific ethics and
accountability in view of the alarming increase of environmental problems.}

If quarks are the actual physical constituents of the neutron, there is no possibility whatever
to utilize the trapped energy, trivially, because the
hypothetical quarks are assumed (but not proved) to be perennially confined inside the neutron.

The neutrino conjecture is even more insidious for environmental issues because it was conceived to deny
Rutherford's conception of the neutron, that is, to deny that the electron is an actual physical
constituent of the neutron. Consequently, there is no known mechanism for tapping the large clean energy
inside the neutron if the neutrino is admitted to be  a physical particle.

On the contrary, if neutrino and quark conjectures are abandoned in favor of the more realistic
view that, being permanently stable, the electron cannot "disappear" at the time of the synthesis of the
neutron and it  is indeed a physical constituent of the neutron, then   there exist  various
 ways for the industrial utilization of the large energy inside the neutron.

Recall that, unlike the proton, the neutron is {\it naturally unstable.} Consequently, it must admit a
{\it stimulated decay.} That predicted by hadronic mechanics was first proposed by Santilli [29] in  1994
and it is given by hitting a selected number of nuclear isotopes, called {\it hadronic fuels,} with hard
photons having a frequency given by a submultiple of the difference of energy between the neutron and the
proton
$$
m_n - m_p = 1.293 MeV = h\times \nu_{reson},
\eqno(5.1a)
$$
$$
\nu_{reson} = (3.288 x 10^{20} Hz) / 1,\; \; \; (3.288 x 10^{20} Hz) / 2,\; \; \; (3.288 x 10^{20} Hz) / 3,\;
\; \; .............
\eqno(5.1b)
$$
under which the isoelectron is expelled by the neutron, resulting in decay
$$
\gamma_{reson} + n \rightarrow p^+ + e^-.
\eqno(5.2)
$$


\begin{figure}
\begin{center}
\epsfxsize=12cm
\parbox{\epsfxsize}{\epsffile{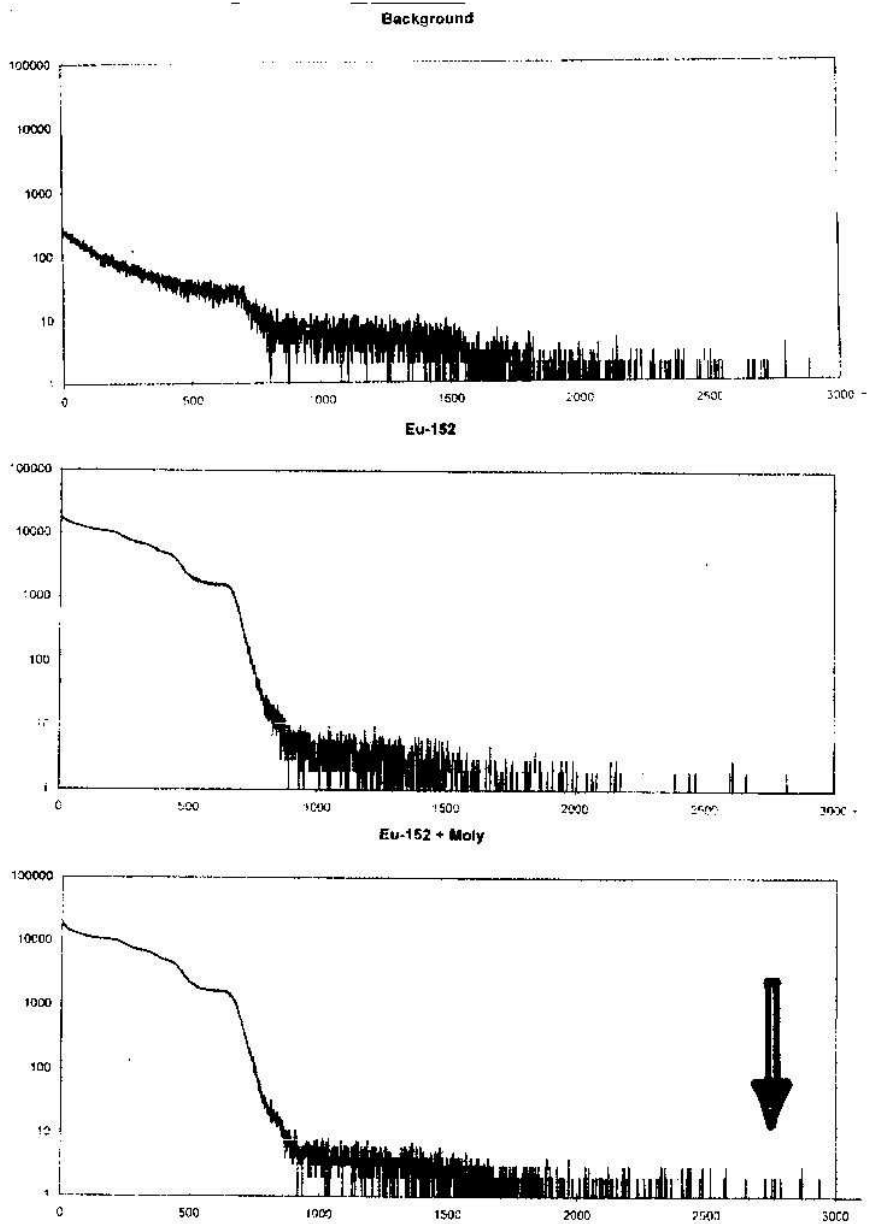}}
\caption{\small \it PRINTOUTS FROM TSAGAS TEST [32]. The top view is a printout of the background; the middle
view is a printout with only the  source of photons with a resonating frequence (a disk of $Eu^156$; and the
bottom view is a printout of the pairing of the Europa source with a disk of Mo(100, 42) showing new energy
lines precisely as predicted by hadronic mechanics for the stimulated decay of the neutron. Unfortunately,
it has not been possible to repeat Tsagas experiment since its original run of 1996 despite numerous
proposals to laboratories around the world because of academic obstructions  dismissing the test on
grounds that the stimulated decay of the neutron is not admitted by neutrino and quark conjectures with
evident damage to environmental research and to society.}
\label{Fig7}
\end{center}
\end{figure}


The energy gain is beyond scientific doubt, because the use of 1/10-th of the exact resonating
frequency (5.1b) would guaranteed the production of a ten-fold clean energy gain. Note that the energy of
the photons not causing stimulated decay is not lost, because absorbed by the hadronic fuel, thus being part
of the heat balance (see Figure 5).

The synthesis of the neutron from protons and electrons was tested experimentally by the Italian
priest-physicist Don Borghi and his associates [31] with positive, yet preliminary results in need of
independent verification. The test can today be confirmed or denied in a variety of way, such as by hitting
a mass of palladium saturated with hydrogen with an electron beam having the threshold energy of $0.782 MeV$
under certain polarizations to assure that the coupling of the electrons and the protons is in singlet as
requested by Figure 4. The detection of neutrons emitted by the mass would confirm synthesis (1.1) at low
energy, thus establishing the validity of hadronic over quantum mechanics beyond credible or otherwise
scientific doubt.

Stimulated decay (5.2) has also been subjected to experimental verifications by Tsagas and his
collaborators [32] with encouraging results also in need of independent verifications. The test was done via
the use of resonating photons originating from a radioactive source hitting a particular isotope of the
molybdenum (an admitted hadronic fuel), according to the reactions
$$
\gamma_{reson} + Mo(100, 42)\rightarrow Tc(100, 43) + \beta^-,
\eqno(5.3a)
$$
$$
Tc(100, 43)\rightarrow Ru(100, 44) + \beta^-,
\eqno(5.3b)
$$
where the first beta decay is stimulated while the second is natural and occurs in 18 sec.

Note that nuclear
energy is based on the disintegration of heavy nuclei, thus implying dangerous radiations and dangerous
waste. By comparison, the hadronic energy is based on the use of {\it light, natural and stable nuclei} as
in case (5.3), thus implying no harmful radiation and no harmful waste because both the original nucleus
Mo(100, 42) and the final one Ru(100,44) are light, natural and stable elements.
For additional detailed studies of the topic, interested readers should consult Refs. [16,29].

To conclude, the condition of contemporary research can be best illustrated by the fact that, following the
expenditure of over one billion dollars resulted in the multiplication of the controversies rather than the
resolution of any of them,  hundred of
millions of dollars continue to be spent by various governments around the world for additional experiments
on neutrino and quark conjectures despite their known catastrophic inconsistencies.

By contrast,
manifestly more important and dramatically less expensive experiments, such as Don Borghi's synthesis of the
neutron [31], Tsagas tests on its stimulated decay [32] and numerous others [14,16], continue to be
discredited, let alone funded and conducted, via the use of neutrino and quark conjectures. In any
case, the investments of various governments on serious environmental research continue to be insignificant,
thus identifying serious problems in our contemporary societies.

The predictable conclusion is that theoretical and experimental research on neutrino
and quark conjectures should indeed continued to be funded, but complemented with the funding of theoretical
and experimental studies on alternative theories without the neutrino and quark conjectures
predicting new clean energies, the latter funding being recommendable not only on grounds of scientific
ethics and accountability, but also to prevent a predictable severe condemnation by posterity, particularly
vis-a-vis the alarming increase of cataclismic climactic events no responsible person can any more deny.

\end{document}